\documentclass[journal,comsoc]{IEEEtran} 
\usepackage{cite}
\usepackage{amsmath,amssymb,amsfonts}
\usepackage{graphicx}
\usepackage{array}
\usepackage{booktabs}
\usepackage{url}
\usepackage{xcolor}
\usepackage[nolist]{acronym}
\usepackage[font=footnotesize,labelfont=bf]{caption}
\usepackage{subcaption}
\usepackage{bm}

\begin{document}
	\begin{acronym}
		\acro{THz}{terahertz}
		\acro{6G}{sixth generation}
		\acro{AP}{access point}
		\acro{RIS}{reconfigurable intelligent surface}
		\acro{ELAA}{extremely large-Scale antenna array}  
		\acro{NLoS}{non-line-of-sight}
		\acro{ISAC}{integrated sensing and communication}
		\acro{SNR}{signal-to-noise ratio}
		\acro{IRS}{intelligent reflecting surface}
		\acro{LoS}{line-of-sight} 
		\acro{ULAA}{ultra-large-scale antenna array}
		\acro{MA}{movable antenna}
		\acro{SWIPT}{simultaneous wireless information and power transfer}
		\acro{AI}{artificial intelligence}
	\end{acronym}
	\title{Bending Beam for THz Wireless Networks: Fundamental, Design Issue, and Prototype}
	\author{Aoran Liu, Xin Tong Hu, Weidong Mei,~\IEEEmembership{Member,~IEEE}, Ya Fei Wu,~\IEEEmembership{Member,~IEEE}, \\Ke Liu, Boyu Ning, Zhi Chen, \IEEEmembership{Senior Member,~IEEE}, and Rui Zhang, \IEEEmembership{Fellow,~IEEE}
		\thanks{A. Liu, W. Mei, K. Liu, B. Ning, and Z. Chen are with the National Key Laboratory of Wireless Communications, University of Electronic Science and Technology of China, Chengdu 611731, China. 
        X. T. Hu and Y. F. Wu are with the School of Electronic Science and Engineering, University of Electronic Science and Technology of China, Chengdu 611731, China. 
        R. Zhang is with the Department of Electrical and Computer Engineering, National University of Singapore, Singapore 117583.}}
	\maketitle
	\begin{abstract}
		\par Bending beams, characterized by their non-diffracting and self-healing properties in the near field, offer a new approach to bypass blockage in \ac{THz} wireless communication and sensing. However, the investigations of bending beams in the context of wireless communications still remain at an early stage. This article provides a state-of-the-art review of the fundamentals and key application scenarios of bending beams in \ac{THz} wireless communications and sensing. We first present and compare the existing beamforming design and practical hardware implementation methods for bending beams. Next, we discuss potential applications of bending beams in wireless communications and sensing and identify their associated challenges, such as blocked channel modeling, bending beam training, codebook design, etc. Finally, a hardware demonstration of bending beam over \ac{THz} frequency bands is presented, validating the advantages of bending beam over conventional beamfocusing.
	\end{abstract}
	
	\section{Introduction}
	\par As a key enabler of \acf{6G} wireless networks, \acf{THz} systems are envisioned to support unprecedented data rates and ultra-high spatial resolution for wireless communications and sensing\cite{ning2023thz}. Particularly, owing to the extremely short wavelength at \ac{THz} frequencies, ultra-large-scale antenna arrays can be densely deployed within compact physical apertures, thereby extending the near-field region and unlocking new spatial degrees of freedom for advanced beam designs such as distance-aware beams~\cite{ning2023thz}. Despite these advantages, \ac{THz} signals are inherently vulnerable to blockages in \ac{LoS} directions, thus severely limiting the link robustness and signal coverage.
	\par
\begin{figure}[!t]
    	\centering
    		\centering   		\includegraphics[width=0.4\textwidth]{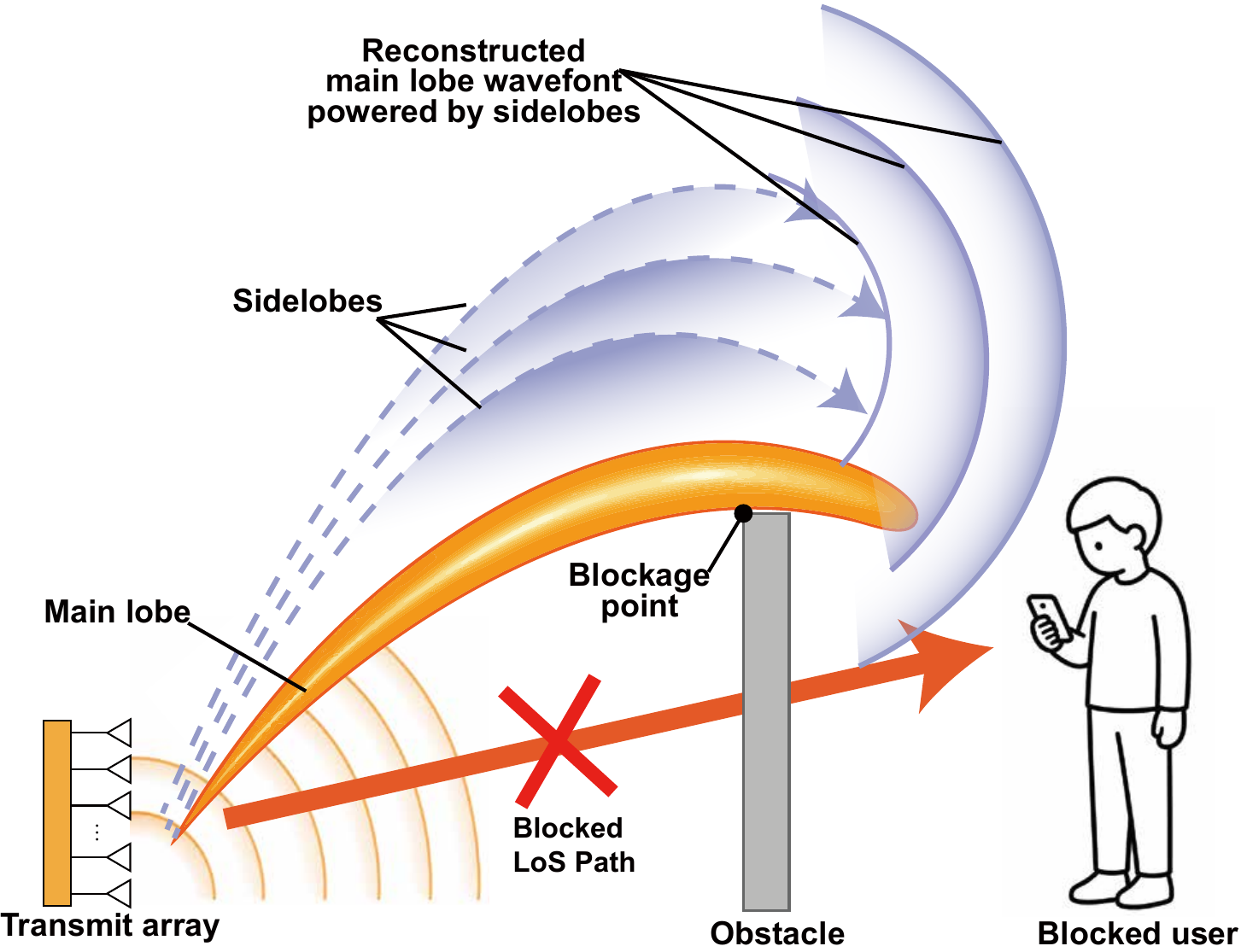}
    		\caption{Physical mechanism of obstacle bypassing via a bending beam.}
    \label{fig:blockage_bypassing_mechanism}
    \end{figure}

 Motivated by recent advances in optics, the concept of bending beams has attracted increasing attention in wireless communications due to their capability to maintain reliable communications even in the presence of LoS blockages~\cite{bending_beam_6G}. Bending beams are typically realized using Airy beams, which are widely recognized in optics for their self-accelerating, non-diffracting nature, enabling curved propagation while maintaining a stable main-lobe profile~\cite{Nonspreading}. As illustrated in Fig.~\ref{fig:blockage_bypassing_mechanism}, conventional near-field beamfocusing concentrates energy on a single target position and exhibits high sensitivity to blockages. In contrast, bending beams can naturally bypass obstacles of certain sizes through their curved propagation and even reconstruct their main-lobe beyond partial obstructions by carefully engineering the amplitude and phase profile of the transmit array. These unique propagation features make bending beams particularly appealing for enhancing link robustness in blockage-prone wireless environments. Furthermore, compared with other anti-blockage technologies such as \ac{IRS}\cite{9771079}, bending beams require no additional infrastructure and leverage direct wavefront engineering at the transmitter, thus offering greater deployment flexibility~\cite{PhysRevApplied}. They can also be synergized with \ac{IRS} to form hybrid curved and reflected propagation paths, thereby further extending coverage and enhancing system resilience.
	\par Despite these promising advantages, most of the existing studies on bending beams have been conducted from optics and electromagnetic perspectives, with primary emphasis on field propagation characteristics and hardware implementation. In contrast, in-depth investigations of bending beams in the context of wireless networks, including their integration with existing wireless technologies, remain at a relatively early stage. In particular, it remains unclear how conventional beamforming and precoding frameworks can be extended to enable effective bending beam control, while simultaneously supporting essential wireless functionalities such as efficient data transmission, reliable channel acquisition, and accurate sensing. 

	\par To facilitate the application of bending beams in \ac{THz} wireless networks, this article provides a state-of-the-art review of their fundamentals and associated key challenges. We begin by reviewing typical wavefront design and hardware implementation methods. Next, we discuss prospective applications of bending beams in wireless communication and sensing systems, such as coverage extension, multi-beam communications, and blocked target sensing. We then identify key challenges and future research directions. Finally, a hardware prototype demonstration is presented to validate the efficacy of bending beams in enhancing THz anti‑blockage communication performance.
	\section{Fundamentals of Bending Beam}
	\par In this section, we introduce the fundamentals of the bending beam, including its physical mechanism for bypassing blockage, beamforming design, and hardware implementation methods. \vspace{-6pt}
    
    \subsection{Physical Mechanism of \ac{LoS} Blockage Bypassing}\label{mechanism}
    \par The physical mechanism underlying bending beams can be understood by examining how their main-lobe energy is formed. For conventional Gaussian (i.e., focused) beams, the main-lobe energy travels predominantly along the LoS link from the transmitter to the receiver, e.g., the blocked user in Fig.\,\ref{fig:blockage_bypassing_mechanism}. When the LoS paths between all transmit antennas and the user are blocked by an obstacle, this dominant energy channel is severely interrupted, resulting in a substantial drop in the field strength behind the obstacle. In contrast, bending beams are designed such that their carried energy mainly propagates along a curved trajectory that can circumvent the blockage, rather than through the \ac{LoS} path, as shown in Fig.~\ref{fig:blockage_bypassing_mechanism}. Even when this curved trajectory is partially intercepted at the obstacle's edge, e.g., the blockage point shown in Fig.~\ref{fig:blockage_bypassing_mechanism}, the Huygens–Fresnel principle comes into play. The unblocked sidelobe components, which still carry the bending‑beam phase information, give rise to secondary sources at the edge. By properly designing the beamforming weights (to be specified in the subsequent subsection), these sidelobe contributions can undergo coherent superposition and thereby reconstruct the main lobe behind the obstacle. In this regard, bending beams are not restricted to THz frequencies, but are generally applicable across the electromagnetic spectrum. At lower frequencies, however, electromagnetic waves usually exhibit stronger diffraction and penetration capabilities, making blockage issues less critical in many practical scenarios. 

   It is also worth noting that there exists a design trade‑off between employing a conventional Gaussian/focused beam and a bending beam, which depends on the severity and geometry of the blockage. Specifically, if a subset of antennas still maintains unobstructed LoS links to the user, that subset could either (i) participate in the bending‑beam generation together with the blocked antennas to improve the overall bending performance, or (ii) form a separate focused beam to maximize the received signal power from the unobstructed paths. The choice between these two strategies involves a compromise between the achievable power gain and the robustness to blockage. 

	\subsection{Bending Beam Design Methods}
	\par
\begin{figure*}[t]
		\centering
		\includegraphics[width=0.75\textwidth]{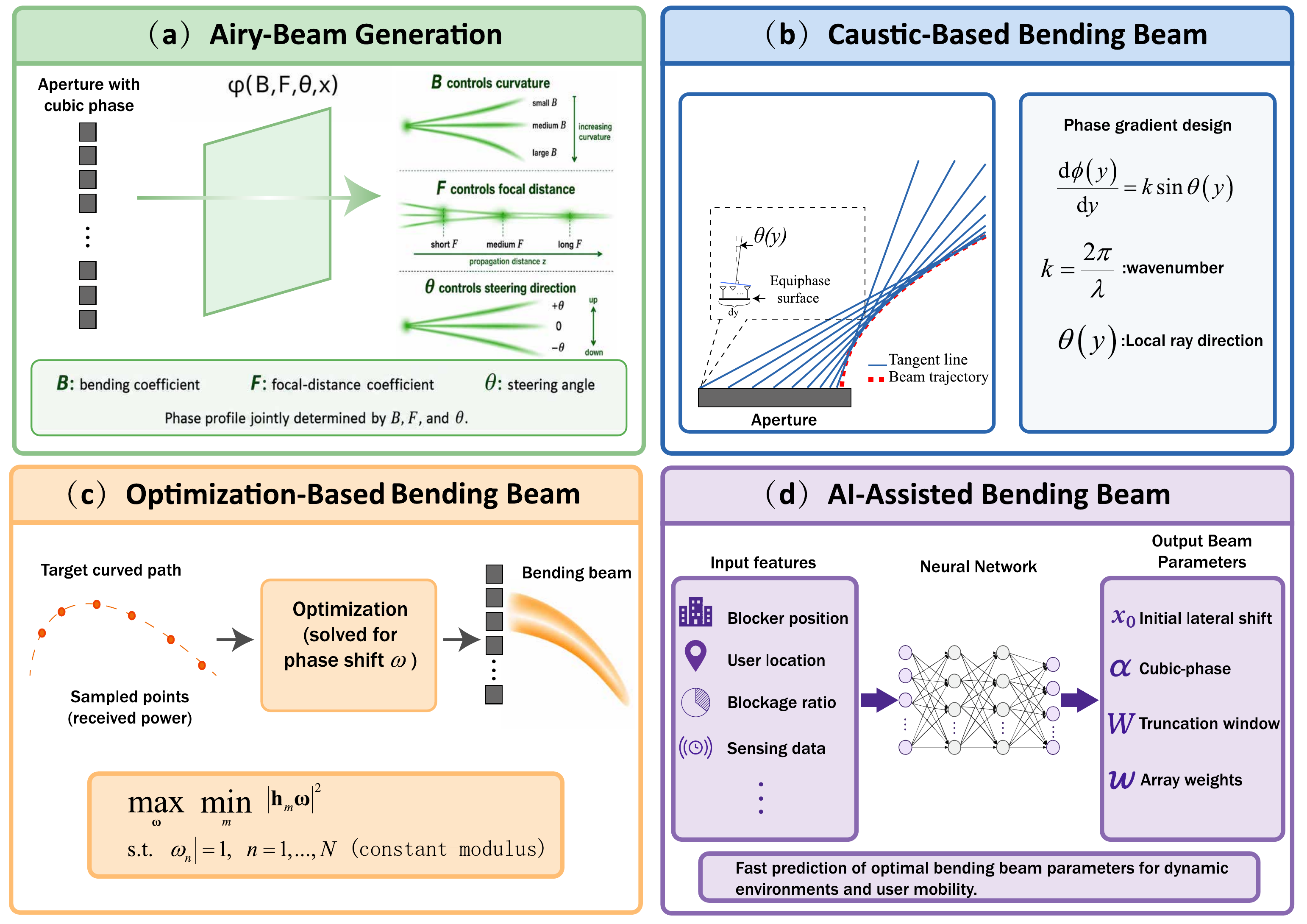}
		\caption{Generation and implementations of bending beam.}
        \vspace{-6pt}
		\label{fig:generate}
	\end{figure*}

 Bending beam requires the design of amplitude and phase profile across the aperture to synthesize wavefronts whose field maxima follow prescribed curved trajectories during propagation. As shown in Fig.~\ref{fig:generate}, existing methods can be broadly grouped into four categories: traditional Airy beams, caustic-based methods, optimization-based methods, and \ac{AI}-driven methods, as discussed next.
    
	\subsubsection{Traditional Airy Beams}
	\par Traditional Airy beams represent the earliest and most fundamental example of bending beams. They provide a basic starting point for understanding how a wave can curve as it propagates. Airy beams were originally introduced as a special class of solutions to the paraxial wave equation~\cite{Nonspreading}, which describes waves propagating mostly along one direction. The most remarkable property of this solution is that its main intensity peak follows a curved trajectory while maintaining a concentrated shape over a considerable distance, instead of spreading out rapidly as other types of beams do (i.e., diffraction). 
	\par As shown in Fig.~\ref{fig:generate}(a), one way to generate an Airy beam is to introduce a non-linear (typically cubic) phase distribution across the aperture. This phase profile causes the radiated wave components to interfere constructively in a coordinated way as they propagate, giving rise to the characteristic curved trajectory. For convenient design and analysis, recent works \cite{zhao2026airy} and \cite{zhang2026breaking} have derived an approximate Airy phase profile in a simple closed-form. In this formulation, the Airy beam can be fully characterized by three parameters, i.e., initial launching angle, which determines the overall beam direction; bending factor, which controls the curvature of the propagation trajectory; and transverse displacement parameter, which specifies the lateral position of the beam caustic.
    
	\subsubsection{Caustic-Based Bending Beams}
	\par While traditional Airy beams serve as the earliest analytical prototype, they can only follow a specific class of beam trajectory (i.e., Airy trajectory), which limits their generality. Caustic‑based methods overcome this limitation by enabling bending beams with a much wider range of convex trajectories. From the perspective of geometric optics, a bending beam can be interpreted as the envelope formed by a family of electromagnetic rays~\cite{10501352}. The basic idea is to treat the radiation emitted from the source aperture, such as an antenna array, as a collection of individual rays. By tailoring the local phase gradient/difference across the aperture, these rays become tangent to a prescribed curved trajectory. The spatial path where the tangent rays accumulate gives rise to the desired bending beam, as illustrated in Fig.~\ref{fig:generate}(b).
    \par Based on this principle, the phase distribution required for a given trajectory is related to the local slope of the curve. To obtain tractable expressions, many studies adopt the paraxial approximation, under which the equivalent rays are assumed to propagate at relatively small angles with respect to the main forward axis. Under this approximation, the required phase profile can often be derived in closed or semi-closed form~\cite{bending_beam_6G}. For the special case of a parabolic beam trajectory, this construction naturally yields the well-known \(3/2\)-power profile\cite{cottrell2009direct}.
	\subsubsection{Optimization-Based Bending Beams}
	\par Optimization-based bending beams represent a fundamental shift from analytical or geometric wavefront construction to objective-driven beam synthesis. As shown in Fig.~\ref{fig:generate}(c), unlike traditional Airy beams or caustic-based designs, optimization-based methods directly formulate bending beamforming as a constrained optimization problem. The beamforming amplitudes and/or phases are optimized according to a target performance metric, such as maximizing the minimum received signal power along a prescribed curved propagation path~\cite{11310498}. In this sense, optimization-based bending beams provide a systematic framework that naturally incorporates communication-oriented design objectives.
	\par More importantly, this framework is not restricted to a particular analytical beam family or geometric construction. By adjusting the optimization objective and constraint set, it can be extended to different target trajectories, propagation requirements, and wireless system configurations. However, such formulations are typically non-convex and may become computationally demanding to solve when practical hardware constraints and large-scale array settings are taken into account.
    
	\subsubsection{\ac{AI}-Driven Bending Beams}
	\par \ac{AI}-driven bending beams represent an emerging data-assisted approach for adaptive Airy beam designs in dynamic wireless environments~\cite{chen2025physics}. Compared with optimization-based approaches, which typically rely on explicit propagation models and iterative numerical solvers, \ac{AI}-based methods aim to directly learn the mapping from environmental observations and system states to suitable bending-beam parameters, e.g., array beamforming weights. As shown in Fig.~\ref{fig:generate}(d), sensing information, historical measurements, and contextual features can be jointly exploited to guide the selection or adjustment of key Airy-beam parameters. 
    \par However, AI‑driven bending beams also face several challenges. For example, their performance heavily depends on the quality and diversity of the training dataset, and insufficient or unrepresentative data may lead to poor generalization in unseen environments. Moreover, the learning process typically requires significant offline training, which can be computationally expensive and time‑consuming. Addressing these limitations is essential for making AI‑driven bending beams practical and reliable in future wireless networks.
    
	\subsubsection{Comparison}
	\par The above four categories of bending beam generation methods reflect a clear evolution from analytical beam construction to adaptive wireless-oriented synthesis. Traditional Airy beams provide the earliest physical prototype and offer intuitive insights into self-bending wave propagation. Caustic-based methods extend this concept by enabling a wider class of beam trajectories. Optimization-based methods provide a general problem formulation to design bending beams, at the cost of higher computational complexity. \ac{AI}-driven methods build on this trend by introducing data-assisted and environment-aware adaptation. Therefore, these methods should be viewed as complementary rather than mutually exclusive, with different strengths in physical interpretability, trajectory flexibility, implementation compatibility, and adaptation capability. The main characteristics of these beamforming design methods are summarized in Table~\ref{tab:comparison_generation_methods}.\vspace{-6pt}
	\begin{table*}[t]
		\caption{Comparison of bending beamforming design methods}
		\centering
		\renewcommand{\arraystretch}{1.2}
		\begin{tabular}{m{3cm}|m{3.2cm}|m{3.5cm}|m{3.5cm}}
			\hline
			\textbf{Method} & \textbf{Core Idea} & \textbf{Main Advantages} & \textbf{Main Limitations}\\
			\hline
			Traditional Airy beams 
			& Analytical phase profile derived from a specific wave solution 
			& Clear physical insights and intuitive interpretation of curved-wave propagation 
			& Limited trajectory flexibility and strong dependence on a specific beam family \\
			
			\hline
			Caustic-based bending beams 
			& Trajectory-oriented synthesis based on ray-envelope construction 
			& More flexible control of beam trajectory 
			& Reliance on geometric approximations and continuous-wavefront assumptions \\
			
			\hline
			Optimization-based bending beams 
			& Objective-driven beam synthesis under communication and system constraints 
			& Direct incorporation of performance metrics and practical design constraints 
			& High complexity due to numerical optimization \\
			
			\hline
			\ac{AI}-driven bending beams 
			& Data-driven learning of the mapping from environment states to bending-beam configurations 
			& Fast online adaptation and environment-aware beam control 
			& Dependence on data quality, training strategy, and generalization ability \\
			\hline
		\end{tabular}
		\label{tab:comparison_generation_methods}
	\end{table*}
    
\subsection{Hardware Implementations}
\par The practical implementation of bending beams depends on how the desired amplitude and phase distribution is implemented on a physical hardware platform. Existing approaches can generally be categorized into three representative classes: reflective/transmissive metasurfaces, holographic metasurfaces, and planar antenna arrays.

\subsubsection{Reflective/Transmissive Metasurfaces}
\par Reflective and transmissive metasurfaces provide a flexible platform for controlling the amplitude, phase, and polarization of electromagnetic waves. Unlike conventional phased arrays, they do not require bulky feeding networks across the entire aperture. Instead, they are illuminated by an external source (e.g., a horn antenna). This makes them attractive for realizing large apertures in \ac{THz} bending‑beam systems. Moreover, metasurfaces offer multiple ways to control the phase, such as using propagation delays through dielectric elements, geometric orientations, or chiral structures. However, achieving simultaneous amplitude and phase control often reduces radiation efficiency for reflective and transmissive metasurfaces, as amplitude modulation typically introduces absorption or scattering losses. In addition, their performance is highly sensitive to phase accuracy, and fabrication imperfections can lead to degraded bending beam quality. 

\subsubsection{Holographic Metasurfaces}
\par Holographic metasurfaces provide a different approach from reflective and transmissive metasurfaces. They modulate surface waves launched by an integrated planar feed, converting them into radiated waves. This mechanism inherently supports simultaneous amplitude and phase control while keeping the structure extremely low‑profile. Importantly, it avoids the free‑space illumination distance required by reflective or transmissive metasurfaces~\cite{quanxi}. The ability of holographic metasurfaces to accurately reproduce complex aperture-field distributions makes them well suited for Airy‑function‑based beam synthesis and other customized bending‑beam designs~\cite{ZHANG}. Despite their potential, the practical realization of holographic metasurfaces at higher frequencies is impeded by the intrinsic difficulty of achieving low-loss surface-wave excitation and propagation in the \ac{THz} regime.

\subsubsection{Planar Antenna Arrays}
\par Planar antenna arrays integrated with radio-frequency feed networks remain one of the most mature hardware platforms for beam generation in microwave and millimeter-wave systems. Their main advantages include low-profile integration, low feed loss, and the ability to realize real-time amplitude and phase control. This capability has been exploited to demonstrate one-dimensional bending beams with microstrip patch arrays~\cite{patch}. 
\par Nevertheless, planar antenna arrays also face several inherent limitations. First, the spatial sampling resolution across the aperture is limited by the minimum spacing between adjacent antenna elements. This discrete sampling limits the array's ability to accurately synthesize rapidly varying phase/amplitude distributions. Consequently, generating two‑dimensional (2D) Airy beams that exhibit rapid amplitude and phase oscillations poses a much greater challenge for planar antenna arrays. Second, the feed network needed for independent element‑level control grows rapidly in complexity with array size, particularly for \ac{THz} bands.

\subsubsection{Comparison}
\par The above three implementation methods exhibit distinct tradeoffs in terms of spatial sampling resolution, wavefront controllability, radiation efficiency, implementation complexity, etc. Reflective and transmissive metasurfaces are well-suited for large-aperture bending beams, especially when phase-only wavefront shaping is sufficient. Holographic metasurfaces provide greater flexibility in reproducing complex field distributions while keeping a compact structure. Planar antenna arrays are mature and compatible with existing wireless hardware, providing high radiation efficiency, while their performance can be limited by feed-network complexity and spatial sampling resolution. The main characteristics of these methods are summarized in Table~\ref{tab:hardware_comparison}.
\begin{figure*}[t]
    \centering
    \includegraphics[width=0.8\textwidth]{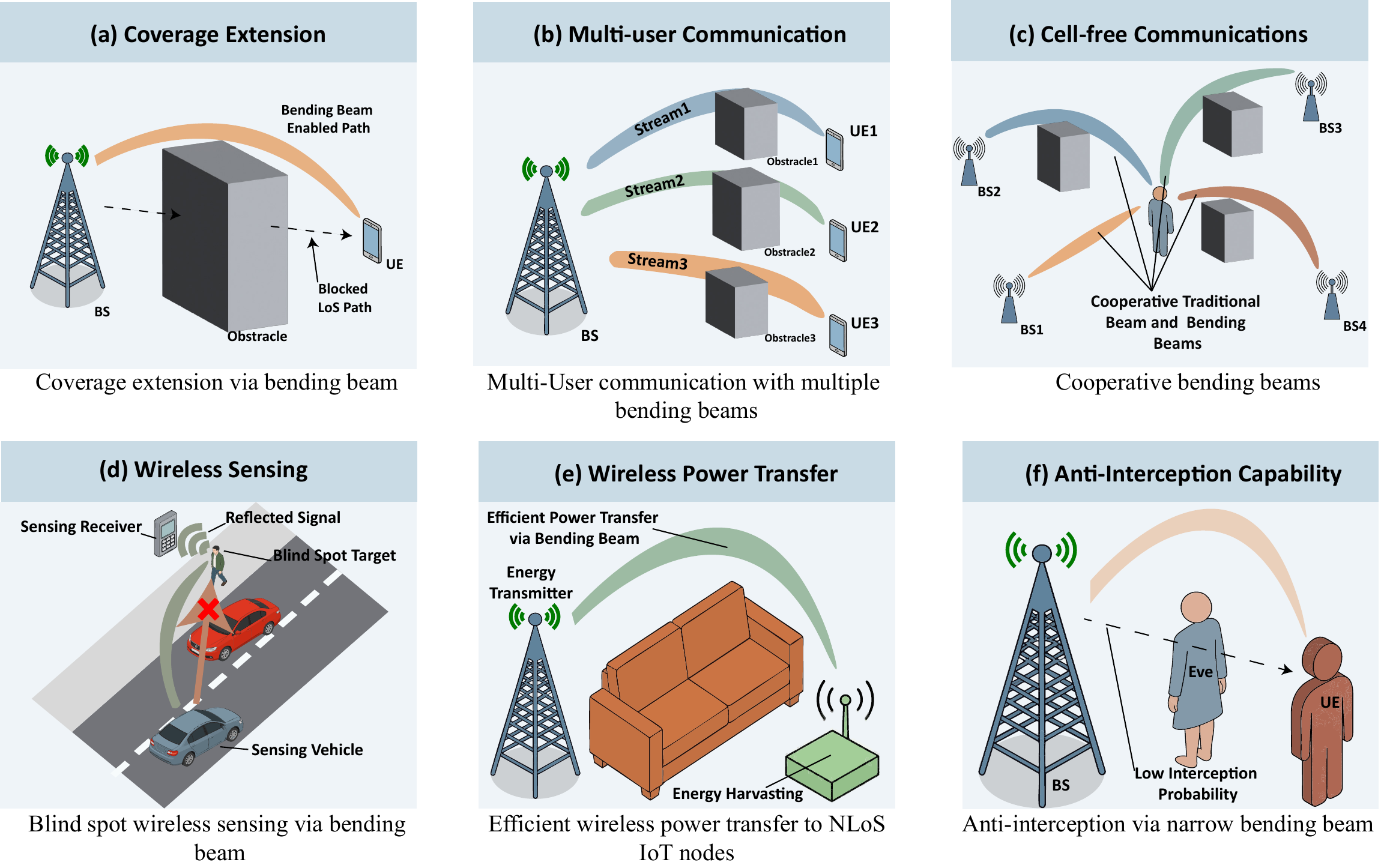}
    \caption{Illustration of typical THz wireless applications using bending beams.}
    \label{fig:intro}
\end{figure*}

\begin{table*}[t]
\centering
\caption{Comparison of hardware implementation methods of bending beams}
\renewcommand{\arraystretch}{1.2}
\begin{tabular}{m{3.1cm}|m{3.5cm}|m{3.6cm}|m{3.8cm}}
\hline
\textbf{Hardware Platform} & \textbf{Core Idea} & \textbf{Main Advantages} & \textbf{Main Limitations} \\
\hline
Reflective and transmissive metasurfaces
& Phase-based wavefront shaping of reflected or transmitted waves
& Large effective aperture, high spatial sampling resolution, and simple passive structure
& Limited real‑time tunability, low radiation efficiency with amplitude control, and high sensitivity to phase accuracy \\
\hline
Holographic metasurfaces
& Joint amplitude‑phase field synthesis using surface waves
& Ultra‑low profile and excellent control over complex field distributions
& Limited real‑time tunability, difficulty in surface‑wave excitation at high frequencies \\
\hline
Planar antenna arrays
& Element‑level amplitude‑phase control via integrated feed network
& High radiation efficiency, mature technology, and real-time tunability
& Feed network complexity and limited spatial sampling resolution \\
\hline
\end{tabular}
	\label{tab:hardware_comparison}
\end{table*}

	\section{Prospective Applications of Bending Beams in Wireless Networks}
	\par In this section, we discuss potential applications of bending beams in wireless networks, covering blockage-aware communication, network coverage enhancement, sensing, energy transfer, and secure communications, etc.\vspace{-6pt}
	
	\subsection{Coverage Extension}
	\par Bending beams can guide electromagnetic energy along curved propagation trajectories, thereby bypassing obstacles and delivering useful signal power to regions that are inaccessible to straight-line beams. This capability is particularly attractive for coverage extension in dense indoor environments, such as offices, factories, and \ac{AI} data centers, where furniture, walls, cabinets, mobile devices, or human activity may interrupt direct links, as shown in Fig.~\ref{fig:intro}(a). By exploiting the additional wavefront control provided by near-field large-aperture arrays, bending beams can improve communication continuity and reduce coverage holes without relying solely on environmental reflections or additional passive infrastructure.\vspace{-6pt}
    
	\subsection{Multi-Beam Communications}
	\par As shown in Fig.~\ref{fig:intro}(b), bending beams also enable a new transmission paradigm for multi-beam communications to serve users with heterogeneous propagation conditions at the same time. In practical scenarios, some users may maintain clear \ac{LoS} links to the transmitter, while others may be located in shadowed or blocked regions due to walls, furniture, vehicles, or human blockage. Conventional beamforming strategies typically struggle to support such heterogeneous users efficiently, since a single straight-line focusing mechanism cannot simultaneously match all propagation conditions.
	\par By exploiting the flexibility of trajectory-aware wavefront engineering, a transmitter can synthesize both conventional focusing beams and bending beams over the same aperture. In this way, \ac{LoS} users can be served by highly focused beams to maximize array gain, while blocked or \ac{NLoS} users can be supported by bending beams whose curved trajectories bypass obstacles and reach shadowed areas. As such, the bending beam extends conventional multi-beam transmission from direction-domain multiplexing to a more general trajectory-domain multiplexing framework. \vspace{-6pt}
    
	\subsection{Cell-Free Communications}
	\par As shown in Fig.~\ref{fig:intro}(c), bending beams are also attractive for cell-free wireless architectures, where multiple distributed access points jointly serve users without rigid cell boundaries. In particular, bending beams provide an additional mechanism for distributed access points to flexibly shape their beam paths according to local blockage conditions. Even when the direct path from one access point to a user is blocked, a bending beam may still reach the user through a designed curved trajectory. Moreover, different access points can generate distinct beam trajectories to jointly support the same user or to serve multiple users. As a result, the effective service region of access points can be significantly enlarged.
	\par More importantly, the distributed access points in a cell-free architecture can be regarded as forming a virtual antenna array with an enlarged spatial aperture. This virtual array strengthens the near-field propagation characteristics and provides additional spatial degrees of freedom for wavefront shaping. Hence, bending-beam transmission becomes even more advantageous compared with conventional single-access-point scenarios.\vspace{-6pt}
    
	\subsection{Wireless Sensing}
	\par As shown in Fig.~\ref{fig:intro}(d), bending beams can also enhance sensing capability in scenarios where targets are located behind obstacles and cannot be directly illuminated by conventional directional beams. Owing to the strong directivity and weak diffraction of \ac{THz} signals, traditional sensing approaches often experience severe performance degradation once the \ac{LoS} path is blocked, leading to blind zones and unreliable echo returns. In contrast, a bending beam can be deliberately engineered to propagate along a curved trajectory that bypasses the obstruction for improved target sensing.
	\par Furthermore, thanks to the reliable communication and sensing enabled by bending beams, they are also highly relevant to emerging \ac{ISAC} applications. For example, multiple bending beams may be generated simultaneously to detect blocked targets and communicate with blocked users. Furthermore, a set of bending beams with different curvatures can be generated to sense the locations and sizes of obstacles. The sensing information can then be leveraged to determine the beam parameters for subsequent communications.\vspace{-6pt}
    
	\subsection{Other Applications}
	\par Another promising application of bending beams is wireless power transfer. As shown in Fig.~\ref{fig:intro}(e), in conventional directional energy transfer, the transmitted power is usually concentrated along a straight beam path, which makes the energy delivery process highly sensitive to blockage and misalignment. Once the direct path is interrupted, the harvested power at the receiver may degrade drastically. In contrast, bending beams can guide electromagnetic energy along curved trajectories, thereby bypassing obstacles and sustaining power delivery to receivers located in partially blocked or hard-to-reach regions. 
	\par Bending beam transmission also offers a new opportunity to improve anti-interception capability and physical layer security in \ac{THz} wireless systems. In conventional beamforming, the transmitted energy is typically concentrated along a straight propagation path, which may unintentionally expose the signal to unauthorized receivers located along or near the main-lobe direction. In contrast, as shown in Fig.~\ref{fig:intro}(f), bending beams enable the transmitter to steer energy along carefully designed curved trajectories. This reduces signal leakage toward unintended spatial regions where potential eavesdroppers may reside. In this sense, bending beams extend conventional directional protection that secures only specific angles and distances to a more general trajectory-level protection mechanism. \vspace{-6pt}
    
	\section{Open Challenges and Future Directions}
	\par Despite the promising advantages of bending beams, several fundamental challenges remain to be tackled before they can be translated into scalable and reliable wireless systems. In the following, we highlight several key open problems and outline representative future research directions for bending beams.\vspace{-6pt}
    
	\subsection{Near-Field Blocked Channel Modeling and Estimation}
	\par Bending beam transmission critically relies on accurate knowledge of the propagation environment (e.g., the positions of blockages) to effectively bypass obstacles. Unlike conventional near-field beamforming, which relies primarily on angular and distance parameters, bending beam design is inherently sensitive to blockage geometry and three-dimensional environmental layout. Hence, conventional near‑field channel models are generally inadequate for accurately characterizing bending‑beam‑enabled blocked links.
	\par This motivates the development of geometry‑aware channel representations for blocked near‑field environments. Promising directions include continuous‑parameter channel modeling, physics‑based propagation descriptions, neural field representation, etc. However, these geometry‑aware channel models also make channel estimation substantially more challenging. Compared with conventional near‑field sparse channel models, the new formulations typically involve a larger number of parameters, more intricate functional expressions, and stronger coupling among distance, angle, blockage state, and spatial non‑stationarity. As a result, traditional sparse channel estimation techniques originally designed for far‑ or near‑field scenarios with multiple reflection paths become less effective in blockage‑dominated environments. This calls for new estimation methods tailored to fully or partially blocked channels. Alternatively, data‑driven approaches may also be explored to circumvent this challenge by incorporating side information from environmental sensing or measurements. 
    	
	\subsection{Trajectory Feasibility and Beam Selection}
    \par Existing bending-beam designs are based on parabolic or convex trajectories. Nevertheless, the maximum achievable curvature and length of such beam trajectories remain open questions. Moreover, the optimal trajectory shape among all feasible trajectories for a given environment remains to be characterized. Another important challenge for bending beams lies in the adaptive selection among focused, bending, and hybrid beamforming modes. As discussed in Section \ref{mechanism}, focused beamforming may still be preferable when sufficiently unobstructed \ac{LoS} paths exist. For partial blockage or specific blockage geometry, hybrid focused-and-bending beamforming can serve as a promising option, for which the antenna aperture and transmit power allocation between the two beam modes turn out to be a new and non-trivial problem to balance the received signal power and blockage robustness. 
    
    Recently, AI‑assisted approaches have been explored for beam selection~\cite{chen2025physics}, by using data‑driven models to choose between focused and Airy beams and to determine their parameters based on environmental observations. While flexible and fast, these methods often require environment‑specific training data and suffer from limited interpretability. Future research should therefore pursue more principled frameworks that integrate physical insight with lightweight learning.\vspace{-6pt}
    \begin{figure*}[!t]
    \subfloat[Hot-stamping bonding process.]{
    \includegraphics[height=0.16\textheight]{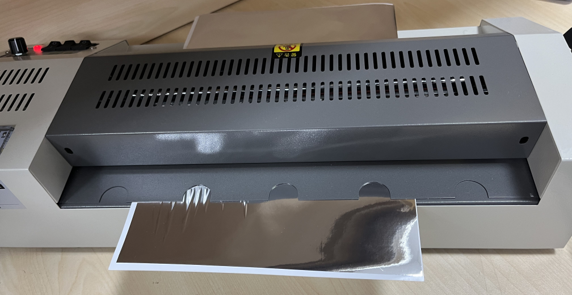}
    \label{fig:fig3_a}
    }
    \subfloat[Measurement platform.]{
    \includegraphics[height=0.16\textheight]{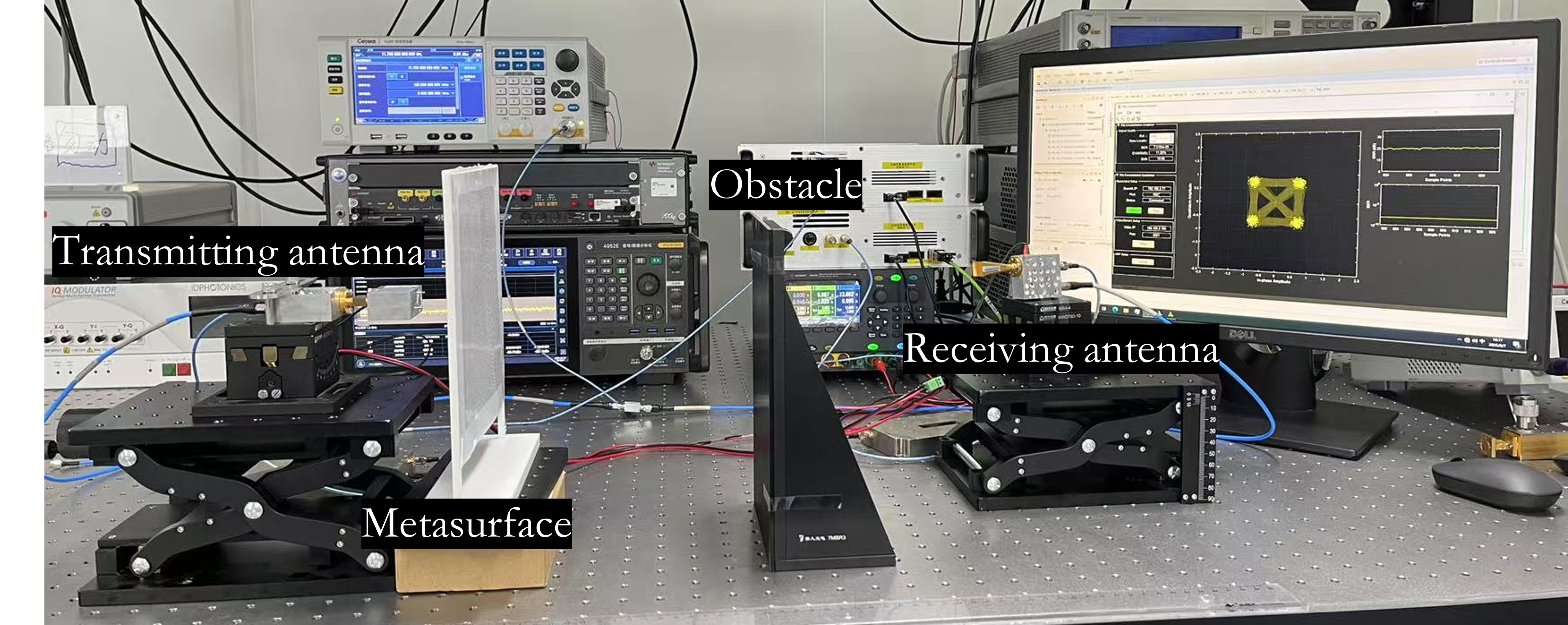}
    \label{fig:fig3_b}
    }\\
    \subfloat[Top view of the propagation geometry.]{
    \includegraphics[height=0.156\textheight]{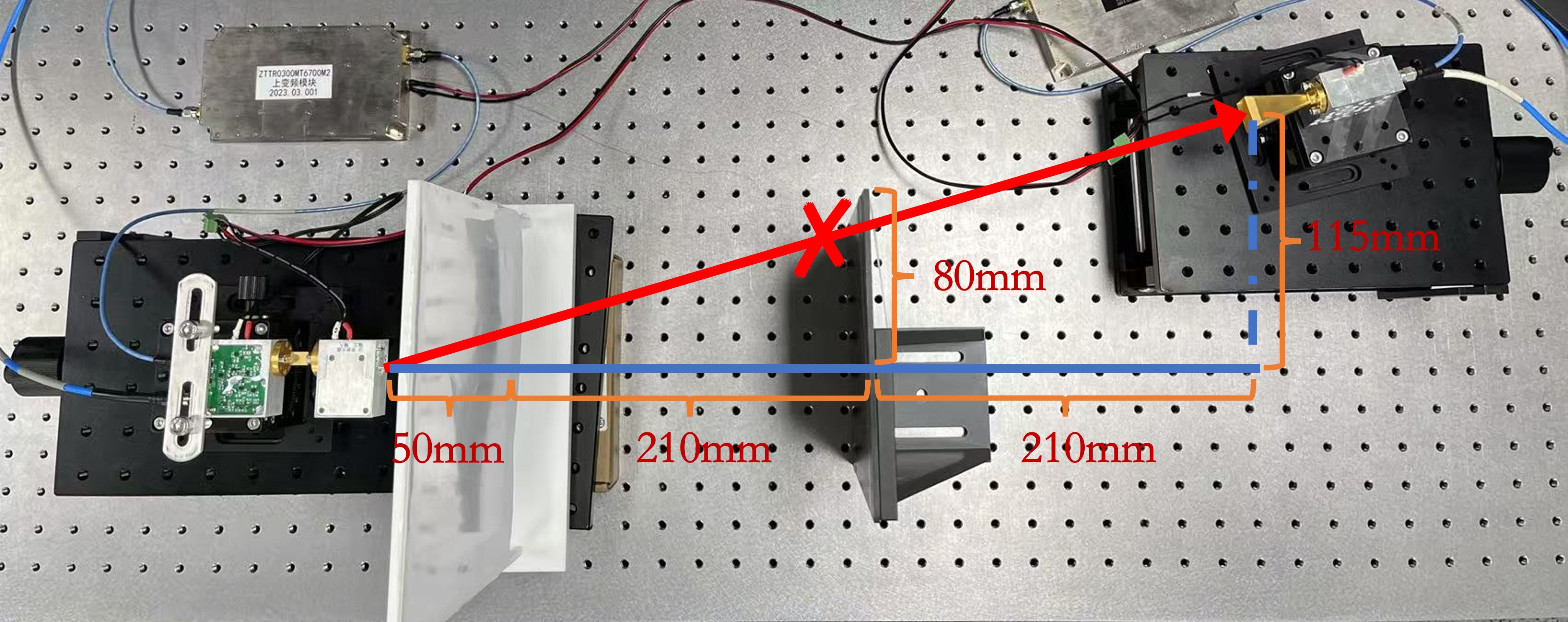}
    \label{fig:fig3_c}
    }
    \subfloat[Measured communication performance under blockage conditions.]{
    \includegraphics[height=0.156\textheight]{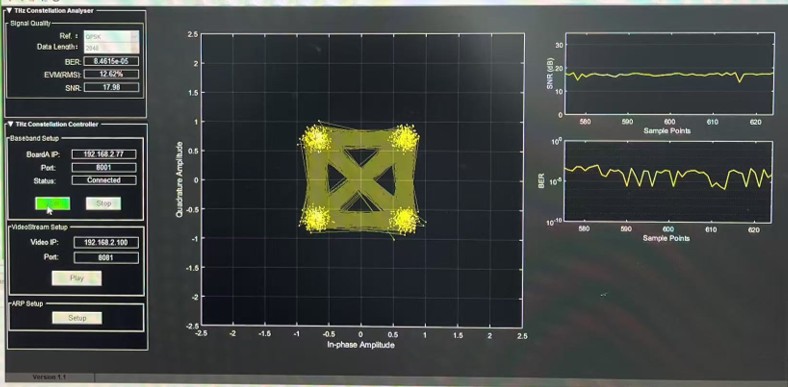}
    \label{fig:fig3_d}
    }
    \caption{Fabrication and experimental results of the metasurface-based bending beam prototype.}
    \label{fig:fig3}
    \vspace{-6pt}
    \end{figure*}
	
	\subsection{Trajectory-Domain Codebook and Beam Training}
	\par Practical deployment of bending beams also requires new beam training and codebook design methodologies. In conventional \ac{THz} systems, beam training is mainly performed over angular codebooks or focal-point codebooks, since the transmitted wave is typically characterized by its dominant direction or focusing location. In contrast, bending beams introduce an additional trajectory dimension into beam design. Therefore, the beam space is no longer limited to angle or focus, but may also involve curvature, launch condition, spatial evolution path, and beam width along the trajectory. 
	\par This enlarged search space makes exhaustive beam training highly inefficient and, in many cases, impractical. In particular, the number of candidate bending-beam patterns may grow rapidly when multiple blockage conditions, user locations, and service objectives must be accounted for jointly. Moreover, trajectory mismatch can be much more detrimental than directional mismatch, since the beam may completely miss the intended blocked region or user location. 
	\par These observations motivate the need for trajectory-domain codebooks and efficient beam training protocols tailored to bending beam systems. Possible directions include hierarchical codebooks, environment-aware beam training, geometry-assisted trajectory initialization, and hybrid codebook structures that jointly support focused beams and bending beams. The tri-parameter characterization of Airy beam proposed in \cite{zhao2026airy} and \cite{zhang2026breaking} also offers an intermediate approach that balances training overhead and performance by restricting the trajectory domain to Airy trajectories. Overall, scalable bending‑beam management strategies will be essential for the practical deployment of bending beams.\vspace{-6pt}
	
	\subsection{Mono-Static Sensing under Path Non-Reciprocity}
	\par Although bending beams provide attraczztive opportunities for blocked target sensing, their application to monostatic sensing remains fundamentally challenging. In conventional monostatic sensing systems based on focused beams, the transmitted and reflected waves approximately follow reciprocal propagation paths, so that the echo can be reliably collected by the same transceiver. However, this basic reciprocity condition does not generally hold for bending beam propagation.
	\par When a bending beam is used to illuminate a blocked or shadowed target, the incident energy is deliberately guided along a curved trajectory. After scattering from the target, the reflected field is not guaranteed to follow the same curved path in the reverse direction. Consequently, the backscattered signal received at the original transmitting node may suffer from severe attenuation, spatial mismatch, or even complete loss. This effect can significantly degrade echo reception quality and thus limit target detection, localization, and parameter estimation performance in monostatic operation. Therefore, it is essential to characterize the propagation properties of bending-beam echoes and properly optimize the deployment of the sensing transmitter accordingly. \vspace{-6pt}
	
	\subsection{Synergy with Other Emerging Wireless Technologies}
	\par The full potential of bending beams may be further unlocked by integrating them with other emerging wireless technologies. For example, bending beams can be combined with \ac{MA} systems \cite{Zhu2026MATutorial}, where adaptive antenna repositioning can enlarge the effective aperture and create more degrees of freedom for curved-wave synthesis. In particular, by flexibly optimizing antenna positions, \ac{MA} systems may achieve high approximation accuracy for the ideal amplitude and phase profiles that demand fine spatial sampling resolution. However, such MA‑enhanced bending beamforming designs also introduce more challenging joint optimization problems involving both antenna positions and beamforming weights, which call for efficient optimization algorithms.
	\par Bending beams may also be integrated with \acp{IRS}\cite{9771079}, enabling hybrid propagation mechanisms that combine direct curved transmission with \ac{IRS} reflection. This approach can further enlarge the coverage region and enhance robustness in highly blocked THz environments. However, allocating power and spatial resources between the IRS‑reflected path and the bending‑beam path becomes a non-trivial problem. The IRS-reflected path offers better controllability but suffers from higher double‑path loss. On the other hand, IRSs may also benefit from monostatic sensing with bending beams by redirecting the echo signal toward the sensing transmitter via an alternative reflected path.\vspace{-6pt}
    
\section{Prototype and Experimental Results}
    \par To show the practical feasibility of bending-beam-enabled \ac{THz} communications, this section presents a metasurface-based bending-beam demonstration operating at 140 GHz. We use a transmissive metasurface in our prototype, fabricated by laser‑printing toner patterns onto paper and then using hot‑stamping thermal transfer to create metal patterns that follow the Airy phase profile (see Fig.~\ref{fig:fig3}(a)). The prototype consists of a transmitting antenna, the phase-engineered metasurface, an intermediate obstacle, and a receiving antenna, as shown in Fig.~\ref{fig:fig3}(b). In particular, an obstacle is deliberately placed between the transmitter and receiver to emulate a blocked \ac{THz} link. The corresponding propagation geometry is depicted in Fig.~\ref{fig:fig3}(c), where the designed curved path bypasses the obstacle without relying on environmental reflections. 
	\par The communication performance of the prototype is shown in Fig.~\ref{fig:fig3}(d). Under the blocked-link condition, communication fails when the metasurface is absent, indicating that the obstructed link cannot be supported by direct transmission alone. By contrast, with the metasurface-enabled self-bending beam, a clear quadrature phase shift keying (QPSK) constellation is observed in Fig.~\ref{fig:fig3}(d), and a wireless transmission rate of 1.6 Gbps is achieved. This clearly verifies the efficacy of the metasurface in reconstructing a viable propagation path around the blockage. More importantly, it shows that bending beams can be implemented with practical hardware to support reliable high-rate \ac{THz} communications in \ac{NLoS} environments.

	\section{Conclusions}
	\par This article has provided an overview of bending beam enhanced wireless communications for future \ac{THz} and \ac{6G} systems. Owing to their self-accelerating, non-diffracting, and self-healing propagation characteristics, bending beams offer a promising way to improve blockage resilience and extend wireless coverage in challenging environments. This article has reviewed representative beamforming design methods, practical implementation methods, and their potential roles in communication, sensing, and other wireless applications.
	\par Despite these advantages, several key challenges remain for bending beams, including accurate near-field blocked channel modeling and estimation, efficient beam training and selection, sensing architecture design, and integration with other emerging technologies. With continued progress, bending beams are expected to become an important enabling technology for future wireless networks. It is hoped that this article serves as a stepping stone for future research on this topic.
	\bibliographystyle{IEEEtran}
	\bibliography{reference} 
\end{document}